\documentclass[aps,prd,reprint,nofootinbib,groupedaddress,longbibliography]{revtex4-2}
\usepackage{amsmath,amssymb}
\usepackage{graphicx}
\usepackage{hyperref}
\hypersetup{
  colorlinks=true,
  linkcolor=blue,
  citecolor=blue,
  urlcolor=blue
}

\newcommand{\fpbh}{f_{\rm PBH}}
\newcommand{\Mmb}{M_{\rm mb}}
\newcommand{\dd}{\mathrm{d}}
\newcommand{\GeV}{\,\mathrm{GeV}}
\newcommand{\g}{\,\mathrm{g}}

\begin{document}

\title{Probing Memory-Burdened Primordial Black Holes with High-Energy Neutrinos}

\author{Arian Moradi Asl}

\author{Sandhya Choubey}

\author{Andreas Lund}

\affiliation{Department of Physics, School of Engineering Sciences, KTH Royal Institute of Technology, AlbaNova University Center, Roslagstullsbacken 21, SE--106~91 Stockholm, Sweden}
\affiliation{The Oskar Klein Centre for Cosmoparticle Physics, AlbaNova University Center, Roslagstullsbacken 21, SE--106 91 Stockholm, Sweden}
\date{\today}

\begin{abstract}
The memory-burden effect can suppress the late-time evaporation of primordial black holes (PBHs), allowing those that would otherwise have evaporated via Hawking radiation to survive until the present epoch. These lighter PBHs emit high-energy and ultra-high-energy neutrinos, opening the tentalizing  possibility of discovery via neutrino telescopes. We study the constraints on memory-burdened PBHs from current IceCube HESE, MESE, and EHE data; and forecast the sensitivity reach of IceCube-Gen2 radio and GRAND200k. In particular, we study how this signal depends on whether the initial PBH population is described by a log-normal mass function or by a monochromatic one. We find that log-normally distributed populations can be more strongly constrained than monochromatic populations with the same median mass primarily because their low-mass tails enhance the high-energy neutrino flux. We show that current IceCube data provide the leading limits for a lower memory burden parameter \(k\), whereas the projected radio detectors become substantially more sensitive for higher values of \(k\). We also study a scenario where future experiments would see a positive signal coming from PBHs. We consider a representative 30-event signal in IceCube-Gen2 radio and GRAND200k and study how well one could distinguish the two mass-function hypotheses. We find that it is easier to disfavor the monochromatic distribution when the log-normal distribution is assumed to be true. Finally, we study how well the parameters of the memory-burdened PBHs can be estimated in these future experiments. 

\end{abstract}

\maketitle

\section{Introduction}
\label{sec:introduction}

PBHs are hypothetical black holes (BHs) that may have formed in the early Universe rather than through the well-established astrophysical formation channels \cite{Zeldovich:1967lct,Hawking:1971}. They can form, for example, from the collapse of large primordial density perturbations, in which case the PBH mass is set by the horizon mass at formation \cite{Carr:1975qj, Carr:1974nx}. Consequently, PBHs serve as probes of small-scale primordial structure and early-Universe dynamics, and may provide an explanation for some, if not all, of the observed dark matter (DM) abundance \cite{Carr:2020xqk, Green:2024bam, Carr:2016drx}.

A promising observational signature of PBHs is Hawking radiation. In the standard semiclassical picture, where one considers quantum fields coupled to classical gravity, a BH emits all kinematically accessible particle species as black-body radiation with temperature \(T_{\rm BH}\propto M^{-1}_{\rm BH}\) \cite{Hawking:1974rv,Page:1976df}. Lighter BHs are therefore hotter, radiate higher-energy particles, and evaporate more rapidly. Unlike stellar BHs, PBHs may form with small enough masses for Hawking radiation to be observable. However, PBHs with initial masses of order \(M\lesssim10^{15}\g\) have already evaporated by the present epoch, whereas heavier PBHs may survive until today \cite{Hawking:1975vcx,Auffinger:2022khh, Carr:2020gox}. Their Hawking radiation nevertheless provides observable particles that can put constraints on the PBH abundance, with recent studies having considered messengers such as gamma rays, cosmic rays, and neutrinos \cite{Dasgupta:2019cae, DelaTorreLuque:2024qms,Bernal:2022swt, Mukhopadhyay:2026lmz}. Furthermore, it has also been investigated whether neutrinos from evaporating PBHs could explain the high-energy events observed by IceCube and KM3NeT \cite{Klipfel:2025jql, Baker:2025cff}.

However, the standard semiclassical description may not capture the full dynamics of BH evaporation, as highlighted by the BH information-loss paradox  \cite{Hawking:1976ra,Raju:2020smc}. 
Recent studies have proposed the memory-burden framework, in which the information stored in an evaporating PBH becomes increasingly costly to maintain as the BH loses mass, thereby suppressing further evaporation \cite{Dvali:2020wft,Dvali:2024hsb}. As a result, this effect may substantially extend the PBH's lifetime, allowing PBHs in mass ranges that would have evaporated by the present epoch to remain viable DM candidates \cite{Alexandre:2024nuo,Thoss:2024hsr, Kohri:2024qpd}.

Memory-burdened PBHs open a wider range of possibilities for detecting high-energy and ultra-high-energy neutrinos emitted via Hawking radiation. For initial masses below the standard survival threshold, memory-burdened PBHs can enter and persist in the burdened phase for astrophysical time scales. Such PBHs can naturally persist into the present epoch, remaining hot enough to emit neutrinos in the TeV-EeV range that propagate over galactic and extragalactic distances with little attenuation. This makes it possible for neutrino telescopes such as the currently running IceCube \cite{IceCube:2016zyt} and KM3NeT \cite{KM3Net:2016zxf} and proposed IceCube-Gen2 \cite{IceCube:2019pna} and GRAND200k \cite{GRAND:2018iaj} to be sensitive to PBHs. 

Since the Hawking temperature is inversely proportional to the PBH mass, lighter PBHs correspond to higher characteristic emission energies. Depending on the strength of the suppression of evaporation, PBHs of different masses may survive until the present epoch, with stronger suppression allowing lighter PBHs to survive and thereby shifting the emitted neutrino spectrum toward higher energies. Because terrestrial neutrino detectors probe different energy ranges, they can test different strengths of the memory-burden suppression. Low energy neutrino experiments such as JUNO and DUNE can constrain the PBH abundance \cite{Bernal:2022swt} in the absence of the memory-burden effect, while stronger suppression can give rise to higher-energy neutrino emission, relevant for other neutrino telescopes. A study constraining memory-burdened PBH parameters using the IceCube EHE \cite{IceCube:2016uab} and 7.5-year HESE \cite{IceCube:2020wum} data, together with projected sensitivities from IceCube-Gen2 radio and GRAND200k, was performed in \cite{Chianese:2024rsn}. 

In this work, we extend and improve this analysis in several ways. We include, for the first time, constraints coming from the IceCube MESE data \cite{IceCube:2025ewu} on the PBH abundance. Furthermore, existing constraints on memory-burdened PBHs have primarily considered a monochromatic mass function \cite{Chianese:2024rsn, Thoss:2024hsr, Chianese:2025wrk, Dvali:2025ktz, Liu:2025vpz, Chaudhuri:2025asm}. Although this is a useful baseline thanks to its simplicity, it is clearly not an exact description. Realistic PBH formation mechanisms generally predict extended mass functions, and constraints on such populations cannot, in general, be obtained by simply evaluating a monochromatic constraint at a single representative mass  \cite{Kuhnel:2017pwq,Carr:2017jsz}. This is especially relevant in the memory-burden scenario, where PBHs with different initial masses may follow distinct evaporation histories. This motivates the present work, where we compare constraints on the PBH abundance for monochromatic and log-normal initial mass functions using IceCube HESE, MESE, and EHE data, together with projected sensitivities for IceCube-Gen2 radio and GRAND200k. We also perform a Bayesian study to assess whether a future PBH neutrino signal could distinguish between the two mass-function models and reconstruct the corresponding PBH parameters. 

The remainder of this article is organized as follows. In Sec. \ref{sec:signal_model}, we describe PBH evaporation in both the standard Hawking and memory-burden regimes, derive the diffuse neutrino flux, and introduce the monochromatic and log-normal mass functions. In Sec. \ref{sec:methods}, we present the statistical analyses of the current IceCube data and the forecasts for future detectors. The resulting upper limits on the PBH abundance, Bayes factors, and posterior distributions are presented in Sec. \ref{sec:results}. We summarize the main conclusions of the analysis in Sec. \ref{sec:discussion}. Additional details on the Bayesian priors and supplementary
parameter-reconstruction results are provided in the appendices.

\section{PBH evaporation and neutrino fluxes}
\label{sec:signal_model}

\subsection{PBH evaporation and the memory-burden effect}
\label{sec:pbh_evolution}

In this work we consider neutral non-rotating PBHs and use natural units, \(\hbar = c = k_B = 1\). Under these assumptions, the instantaneous mass \(M\) is the only BH parameter governing the temperature and mass evolution of Hawking radiating BHs. The Hawking temperature is given by \cite{Hawking:1975vcx}
\begin{equation} 
\label{eq:HawkingTemperature}
T_{\rm BH} = \frac{1}{8\pi G M} \simeq 1 \left(\frac{10^{13}\g}{M}\right) \GeV ,
\end{equation}
where \(G\) is Newton's gravitational constant. The inverse mass scaling implies that lighter PBHs are hotter and therefore radiate higher-energy particles.

In the standard semiclassical picture, the instantaneous emission rate for a particle species \(j\) is given by \cite{Page:1976df}
\begin{equation}
\frac{\dd^2 N_j(E, M)}{\dd E\,\dd t} = \frac{g_j}{2\pi} \frac{\Gamma_j(E,M)}{\exp(E/T_{\rm BH})\pm1},
\label{eq:hawking_spectrum}
\end{equation}
where \(g_j\) is the number of internal degrees of freedom for species \(j\) and \(\Gamma_j(E,M)\) denotes the greybody factor after summing over angular-momentum modes. The \(-\) (\(+\)) sign applies for bosons (fermions).

As the PBH emits Hawking radiation, the emitted particles carry energy away from the BH, causing its mass to decrease with time. In the semiclassical picture, the rate of this mass loss is expressed as \cite{Mazde:2022sdx}
\begin{align}
\frac{\dd M}{\dd t} &= -\sum_j \frac{g_j}{2\pi} \int_0^\infty \frac{\Gamma_j(E,M)}{\exp(E/T_{\rm BH})\pm1}E \,\dd E \\ &= -\frac{\alpha(M)}{M^2},
\label{eq:standard_mass_loss}
\end{align} 
where
\begin{equation}
\alpha(M) \equiv \frac{\mathcal{G}\,g_H(T_{\rm BH})}{30720\pi G^2}.
\end{equation}
Here, \(\mathcal{G}\simeq3.8\) is an effective BH greybody coefficient due to near-horizon propagation dynamics and \(g_H(T_{\rm BH})\) is the spin-weighted number of degrees of freedom at temperature \(T_{\rm BH}\). For the mass ranges of the PBHs considered in this work, we approximate \(g_H(T_{\rm BH})\) by the constant high-temperature Standard Model (SM) value
\(g_H \simeq 102.6\). For an initial PBH mass \(M_i\), integrating Eq \eqref{eq:standard_mass_loss} yields
\begin{equation}
M(t)=
\left[
M_i^3-3\alpha(t-t_i)
\right]^{1/3}.
\label{eq:standard_mass_evolution}
\end{equation}
The corresponding lifetime of the PBH is then
\begin{equation}
\tau_{\rm PBH}
=
\frac{M_i^3}{3\alpha}
\simeq
4.3\times10^{17}
\left(\frac{M_i}{10^{15}\g}\right)^3
\mathrm{s},
\label{eq:standard_lifetime}
\end{equation}
where \(t_0\simeq4.3\times10^{17}\,{\rm s}\) is approximately the age of the Universe. Thus, Eq.\eqref{eq:standard_lifetime} shows the survival threshold, where PBHs with \(M_i\lesssim10^{15}\g\) have fully evaporated before the present epoch.

The memory-burden framework describes a possible modification of BH evaporation in which the information carried by the PBH is stored in internal degrees of freedom known as memory modes. As the PBH loses mass, maintaining the stored information becomes increasingly energetically costly. This growing memory burden backreacts on the evaporation process, suppressing further evaporation \cite{Dvali:2024hsb}.

We implement this framework phenomenologically as a two-stage modification of the standard Hawking evaporation process, following Refs. \cite{Haque:2024eyh,Chianese:2024rsn}. During the initial phase, the PBH evolves according to the standard Hawking evaporation framework. The subsequent memory-burdened phase, in which evaporation is suppressed begins once the mass of the PBH decreases to a fraction of its initial value
\begin{equation}
M_q=qM_i.
\label{eq:mq_def}
\end{equation}
Throughout the analysis we adopt \(q=1/2\), following previous studies and motivated by the expectation that the backreaction due to memory burden becomes relevant no later than when the PBH has lost half of its initial mass. The corresponding onset time for the memory-burdened phase follows from Eq.\eqref{eq:standard_mass_evolution} and is given by
\begin{equation}
t_q=\tau_{\rm PBH}(1-q^3),
\label{eq:tq_def}
\end{equation}
where \(\tau_{\rm PBH}\) is the lifetime of the PBH. Once the PBH has lost half of its initial mass, the memory-burden effect becomes significant and the instantaneous emission rate is suppressed by a negative power of the BH entropy \(S(M)=4\pi G M^2\) such that
\begin{equation}
\frac{\dd^2N_\nu^{\rm mb}(E_\nu, \Mmb)}{\dd E_\nu\,\dd t} = \frac{1}{S[\Mmb(t)]^k} \frac{\dd^2N_\nu(E_\nu, \Mmb)}{\dd E_\nu\,\dd t}.
\label{eq:mb_emission}
\end{equation}
Here, \(k > 0\) is a parameter that controls the strength of the backreaction due to the memory-burden effect. The limit \(k=0\) recovers standard Hawking evaporation, while larger values lead to stronger suppression that may radically increase the lifetime of the PBH. Consequently, the mass-loss rate of the PBH is suppressed by the same BH entropy factor
\begin{equation}
\frac{\dd \Mmb}{\dd t} = \frac{1}{S[\Mmb(t)]^k} \left.\frac{\dd M}{\dd t}\right|_{M = \Mmb(t)}.
\label{eq:mb_mass_loss}
\end{equation}
Integrating Eq. \eqref{eq:mb_mass_loss} then gives 
\begin{equation}
\Mmb(t) = M_q \left[1-\lambda_{\rm mb}^{(k)}(t-t_q)\right]^{1/(3+2k)}, \qquad t\geq t_q,
\label{eq:mb_mass_evolution}
\end{equation}
where \(\lambda_{\rm mb}^{(k)}\) is the inverse timescale of the memory-burdened phase given by
\begin{equation}
\lambda_{\rm mb}^{(k)} = \frac{(3+2k)\alpha} {(4\pi G)^kM_q^{3+2k}}.
\label{eq:gamma_pbh_k}
\end{equation}
The total lifetime of a memory-burdened PBH is therefore
\begin{equation}
\tau_{\rm PBH}^{(k)} = t_q+ \left(\lambda_{\rm mb}^{(k)}\right)^{-1}.
\label{eq:mb_lifetime}
\end{equation}

The prescription used here corresponds to an instantaneous transition from the initial phase to the memory-burdened phase at \(M = M_q\). More general smooth transitions, in which the suppression turns on gradually rather than discontinuously, have also been considered \cite{Dvali:2025ktz,Montefalcone:2025akm, Dondarini:2025ktz}. In these treatments the PBH can spend a finite period in an intermediate regime between the standard Hawking phase and the fully memory-burdened phase, which can significantly alter its evolution relative to the instantaneous transition framework. In this work, we restrict ourselves to the transition prescription defined above.

\subsection{Diffuse neutrino flux}
\label{sec:diffuse_flux}

The neutrino emission spectrum from a PBH can be decomposed into a primary and a secondary component \cite{Bernal:2022swt},
\begin{equation}
\frac{\dd^2N_\nu(E_\nu,M)}{\dd E_\nu\,\dd t} = \left. \frac{\dd^2N_\nu(E_\nu,M)}{\dd E_\nu\,\dd t}
\right|_{\rm prim} + \left. \frac{\dd^2N_\nu(E_\nu,M)}{\dd E_\nu\,\dd t}
\right|_{\rm sec}.
\label{eq:primary_secondary}
\end{equation}
The primary component corresponds to neutrinos emitted directly through Hawking radiation, while the secondary component arises from the decays of other primary particles and their hadronization products. For the present analysis, the emission spectra are evaluated with \textsc{BlackHawk} v2.3 \cite{Arbey:2019mbc, Arbey:2021mbl} for neutral and non-rotating PBHs, with the secondary neutrino spectra evaluated through \textsc{HDMSpectra} \cite{Bauer:2020jay} \footnote{The \textsc{BlackHawk} output obtained with the \textsc{HDMSpectra} option follows a single-particle convention. We therefore apply the rescaling prescription described in Appendix A of Ref. \cite{Dondarini:2025ktz}.}. In the standard \textsc{BlackHawk} setup, neutrinos are treated as massless with six total degrees of freedom, corresponding to the Majorana neutrino convention. The alternative Dirac case, with \(g_\nu = 12\), would introduce additional degrees of freedom and therefore increase the diffuse neutrino flux from the PBHs \cite{Lunardini:2019zob}. This case is however not considered in the present work. Nevertheless, the analysis uses the diffuse all-flavor flux, summed over active neutrino flavors and antineutrinos. The observed diffuse neutrino flux receives contributions from PBHs in the DM halo of our galaxy as well as from extragalactic PBHs. For a monochromatic PBH population, the galactic flux can be written as \cite{Chianese:2024rsn}
\begin{equation}
\frac{\dd^2\Phi_\nu^{\rm gal}}{\dd E_\nu\,\dd\Omega}(E_\nu) = \frac{\fpbh}{4\pi \Mmb(t_0)}
\frac{\dd^2N_\nu^{\rm mb}(E_\nu,M_{\rm mb}(t_0))}{\dd E_\nu\,\dd t}\mathcal{J}_{\Delta \Omega}, 
\label{eq:galactic_flux}
\end{equation}
where
\begin{equation}
    \mathcal{J}_{\Delta \Omega} = \frac{1}{\Delta \Omega} \int_{\Delta \Omega} d\Omega \int_{0}^\infty d s \rho_{\rm DM}[r(s,\psi)],
\end{equation}
is the J-factor averaged over a solid angle \(\Delta \Omega\). Here \(\Mmb(t_0)\) is the present-day PBH mass and \(\fpbh = \rho_{\rm PBH}/\rho_{\rm DM}\) is the present-day PBH abundance. For the DM density profile, we adopt a Navarro-Frenk-White (NFW) profile \cite{Navarro:1996gj}
\begin{equation}
    \rho_{\rm DM}(r) = \frac{\rho_s}{(r/r_s)(1+r/r_s)^2},
\label{eq:nfw_profile}
\end{equation}
where
\begin{equation}
r(s,\psi) = \left(s^2+R_\odot^2-2sR_\odot\cos\psi\right)^{1/2},
\label{eq:galactocentric_dist}
\end{equation}
is the galactocentric distance along a line of sight. Here, \(s\) is the line of sight distance, \(\psi\) is the angular separation from the galactic centre, and \(R_\odot\) is the distance between the Sun and the galactic centre. Following Ref.~\cite{Chianese:2025wrk}, we adopt
\(R_\odot=8.178\,\mathrm{kpc}\), \(r_s=25\,\mathrm{kpc}\), and \(\rho_s=0.23\, \mathrm{GeV\,cm^{-3}}\), corresponding to a local DM density of \(\rho_{\rm DM}(R_\odot)=0.4\, \mathrm{GeV\,cm^{-3}}\). For a full-sky average of the J-factor, we have that
\begin{align}
\mathcal{J}_{4\pi} &= \frac{1}{4\pi} \int \dd\Omega \int \dd s\, \rho_{\rm DM}[r(s,\psi)]  \\ &= 2.22\times10^{22}\, \GeV\,\mathrm{cm}^{-2}.
\label{eq:jfactor_value}
\end{align}

The extragalactic contribution is instead obtained by integrating the redshifted neutrino emission from the cosmological PBH population \cite{Liu:2023cqs},
\begin{equation} 
\frac{\dd^2\Phi_\nu^{\rm eg}(E_\nu)}{\dd E_\nu\ \dd\Omega} = \frac{\fpbh\rho_{\rm DM}}{4\pi\Mmb(t_0)} \int_{0}^{z_{\rm max}}\frac{dz}{H(z)}\, \frac{\dd^2N_\nu^{\rm mb}(E_\nu', M[t(z)])}{\dd E_\nu'\,\dd t},
\label{eq:eg_flux}
\end{equation}
where \(\rho_{\rm DM} \simeq 1.3 \times 10^{-6}\,{\rm GeV\,cm^{-3}}\) is the present-day average DM density of the universe, and
\begin{equation}
    H(z) = H_0\sqrt{\Omega_r(1+z)^4+\Omega_m(1+z)^3+\Omega_\Lambda},
\end{equation}
is the Hubble expansion rate. Here, we adopt \(H_0=67.4\,{\rm km\,s^{-1}\,Mpc^{-1}}\), \(\Omega_m=0.315\), \(\Omega_\Lambda=0.685\), and \(z_{\rm eq}\simeq3400\) \cite{Planck:2018vyg}. The radiation density parameter follows from the matter--radiation equality as \(\Omega_r=\Omega_m/(1+z_{\rm eq})\simeq9.2\times10^{-5}\). In the integrand of Eq. \eqref{eq:eg_flux}, the emission spectrum is evaluated at the emitted neutrino energy \(E_\nu' = (1+z)E_\nu\), and at the PBH mass \(M[t(z)]\), where \(t(z)\) is the cosmic time corresponding to redshift \(z\). Furthermore, we take \(z_{\rm max} = z_{\rm eq}\), corresponding to matter-radiation equality, and have verified that extending the integral beyond equality changes the predicted flux negligibly.

Therefore, the total sky-averaged all-flavor diffuse neutrino flux from PBHs in the galactic DM halo and the isotropic extragalactic DM distribution is given by

\begin{equation}
    \frac{\dd^2\Phi_\nu^{\rm tot}(E_\nu)}{\dd E_\nu\,\dd\Omega} = \frac{\dd^2\Phi_\nu^{\rm gal}(E_\nu)}{\dd E_\nu\,\dd\Omega} + \frac{\dd^2\Phi_\nu^{\rm eg}(E_\nu)}{\dd E_\nu\,\dd\Omega}.
    \label{eq:total-flux}
\end{equation}

\subsection{Extended mass functions}
\label{sec:mass_functions}

Monochromatic mass functions are commonly used when studying PBHs as they provide a useful reference case. However, well-motivated PBH formation mechanisms do not produce exactly monochromatic populations, but rather yield different types of extended mass functions. For example, PBHs formed from the collapse of large primordial perturbations produces broadened mass distributions via critical collapse \cite{Gorton:2024cdm}. More generally, the shape and width of the peak in the primordial power spectrum also affect the resulting PBH mass function \cite{Gow:2020cou}. Extended mass functions can also arise near the QCD phase transition, where the softening of the equation of state modifies the collapse process and enhances PBH formation over a range of masses \cite{Musco:2023dak}. It is therefore of interest to consider the phenomenological consequences of an extended mass function. 

In this work, we adopt a log-normal initial mass distribution. This distribution should not be interpreted as an exact prediction of any particular formation scenario. Rather, it is used as a standard benchmark for studying how an extended PBH mass function modifies the phenomenology as compared to the monochromatic case.

To implement these two cases, we describe the initial PBH population by the mass function \cite{Gorton:2024cdm}
\begin{equation}
\psi_i(M_i, t_i)
\equiv
\frac{1}{\rho_{\rm PBH}^{i}}
\frac{\dd\rho_{\rm PBH}(M_i, t_i)}{\dd M_i},
\label{eq:mass_function_norm}
\end{equation}
where \(\rho_{\rm PBH}(M_i, t_i)\) is the comoving mass density in PBHs of initial mass \(M_i\) at formation time \(t_i\), and \(\rho_{\rm PBH}^{i}\) is the total initial comoving mass density of PBHs.
For the monochromatic distribution, we take
\begin{equation}
\psi_i^{\rm mono}(M_i, t_i)=\delta(M_i-M_0),
\label{eq:mono_mass_function}
\end{equation}
where \(M_0\) is the initial PBH mass. For the log-normal distribution, we take 
\begin{equation}
\psi_i^{\rm LN}(M_i, t_i)
=
\frac{1}{\sqrt{2\pi}\sigma M_i}
\exp\left[
-\frac{\ln^2(M_i/M_c)}{2\sigma^2}
\right],
\label{eq:lognormal_mass_function}
\end{equation}
where \(M_c\) is the median initial mass and \(\sigma\) controls the width of the distribution.

Since these mass functions are specified in terms of initial masses, we keep \(M_i\) as the integration variable and evolve each initial mass to its present-day value \(\Mmb(t_0)\). However, Hawking evaporation can alter an extended PBH mass function between formation and present-day, primarily at low masses where PBHs may lose a substantial fraction of their initial mass. This must therefore be taken into consideration when constructing the present-day mass function \cite{Mosbech:2022lfg}. We denote the range of initial masses that survive until today by \(\mathcal{S}_k\), defined by \(\tau_{\rm PBH}^{(k)}(M_i) > t_0\). Furthermore, the neutrino emission is evaluated only for PBHs whose instantaneous masses lie within the mass range \(10^{-1}\,{\rm g}\leq \Mmb(t)\leq10^{10}\,{\rm g}\).  The normalized present-day distribution over initial masses is then
\begin{equation}
{\cal W}(M_i)
=
\frac{1}{\mathcal{A}_k}
\psi_i(M_i, t_i)\,\Mmb(t_0)/M_i,
\label{eq:survivor_weight}
\end{equation}
where \(\mathcal{A}_k = \int_{\mathcal{S}_k}\dd M_i'\,
\psi_i(M_i', t_i)\,\Mmb(t_0)/M_i'\) is the fraction of the initial PBH mass density remaining today and the factor \(\Mmb(t_0)/M_i\) accounts for the mass lost between formation and the present epoch. Thus, Eq. \eqref{eq:survivor_weight}
gives the normalized present-day PBH mass-density distribution of the surviving PBHs, expressed in terms of their initial masses. The quantity \(\fpbh{\cal W}(M_i)\dd M_i\) gives the fraction of the present-day DM density in the form of surviving PBHs that formed with initial masses in the interval \([M_i,M_i+\dd M_i]\). With this convention, \(\fpbh = 1\) corresponds to the case where the surviving PBH population accounts for all of the DM today.

Thus, the galactic contribution from a log-normally distributed PBH population is given by
\begin{align}
\frac{\dd^2\Phi_\nu^{\rm LN, gal}(E_\nu)}
{\dd E_\nu\,\dd\Omega}
&=
\fpbh
\int_{\mathcal S_k}\dd M_i\,
{\cal W}(M_i)
\nonumber\\
&\quad\times
\left.
\frac{\dd^2\Phi_\nu^{\rm mono, gal}(E_\nu)}
{\dd E_\nu\,\dd\Omega}
\right|_{M_i,k,\fpbh=1}.
\label{eq:extended_galflux}
\end{align}
Here \(\dd^2\Phi_\nu^{\rm mono, gl}(E_\nu)/(\dd E_\nu\dd\Omega)\) denotes the galactic diffuse neutrino flux from a monochromatic PBH population with initial mass \(M_i\), memory burden parameter \(k\), and unit present-day PBH abundance. The extragalactic contribution from a log-normally distributed PBH population is instead given by
\begin{align}
\frac{\mathrm{d}^2\Phi_{\nu}^{\rm LN, eg}(E_\nu)}
{\mathrm{d}E_\nu\,\mathrm{d}\Omega}
&=
\frac{f_{\rm PBH}\rho_{\rm DM}}
{4\pi \mathcal{A}_k}
\int \mathrm{d}M_i\,
\frac{\psi_i(M_i)}{M_i} \nonumber\\
&\quad\times
\int_{z_{\min}(M_i)}^{z_{\rm eq}}
\frac{\mathrm{d}z}{H(z)}
\frac{\mathrm{d}^2N_\nu^{\rm mb}
\!\left(E_\nu',M[t(z)]\right)}
{\mathrm{d}E_\nu'\,\mathrm{d}t},
\label{eq:extended_egalflux}
\end{align}
where \(z_{\rm min}(M_i) = 0\) if the PBH survives until today, and otherwise the redshift at which a PBH with initial mass $M_i$ evaporates.

\begin{figure}[t]
\centering
\includegraphics[width=\columnwidth]{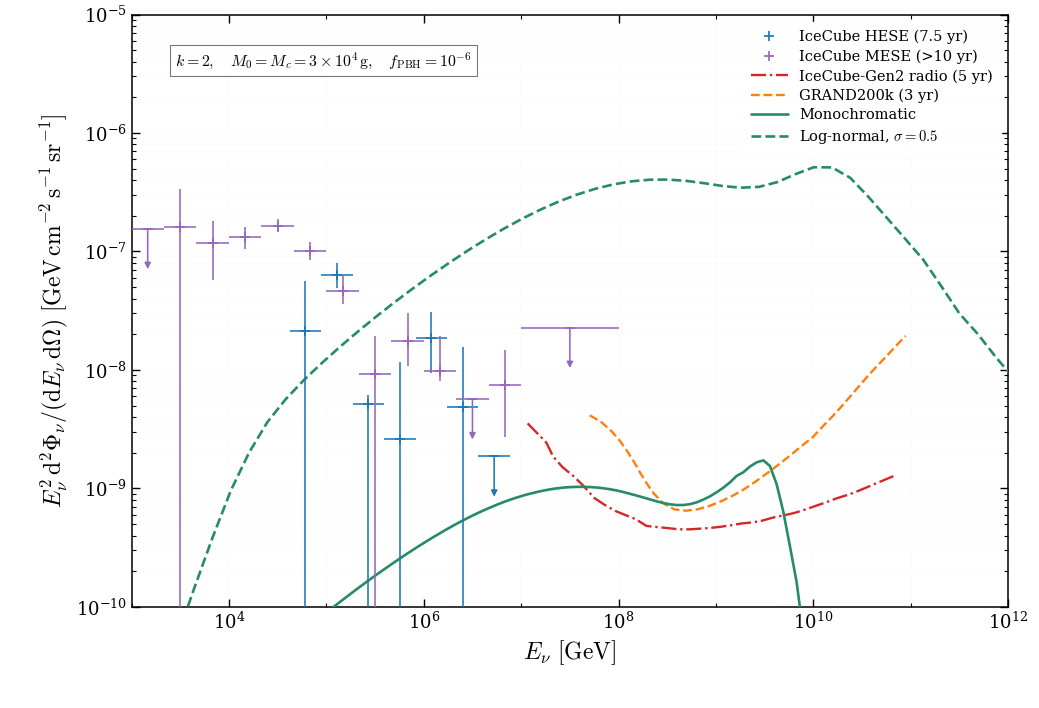}
\caption{Illustrative sky-averaged all-flavor diffuse neutrino flux from memory-burdened PBHs for a set of benchmark parameters. Solid and dashed green curves denote the monochromatic and log-normal PBH spectrum, respectively. Current IceCube HESE \cite{IceCube:2020wum} and MESE \cite{IceCube:2025ewu} segmented flux measurements and published \(90\%\) CL differential sensitivities for IceCube-Gen2 radio \cite{IceCube:2019pna} and GRAND200k \cite{GRAND:2018iaj} are shown for comparison. These sensitivity curves are included only as visual references, and are not used for the forecasts below.}
\label{fig:benchmark-flux-data}
\end{figure}

Fig. \ref{fig:benchmark-flux-data} compares the total sky-averaged all-flavor diffuse neutrino
fluxes for illustrative monochromatic and
log-normal benchmarks with the same memory burden parameter and characteristic initial mass, \(M_0=M_c\). The log-normal spectrum is broader and extends to higher energies because its low-mass tail contains lighter surviving PBHs with larger Hawking temperatures. Consequently, a larger fraction of its flux overlaps with the energy ranges targeted by future radio-neutrino detectors.

\section{Analysis methods}
\label{sec:methods}

\subsection{Analysis of current IceCube data}
\label{sec:current_likelihoods}

We now derive constraints on the present-day abundance of memory-burdened PBHs from three separate IceCube datasets. The HESE \cite{IceCube:2020wum} and MESE \cite{IceCube:2025ewu} datasets are accounted for using their predicted measurements of the segmented diffuse all-flavor flux, while the IceCube EHE analysis is incorporated using its observed event count \cite{IceCube:2016uab}. The HESE and MESE segmented flux measurements used in this work are shown in Fig. \ref{fig:benchmark-flux-data}. We derive a separate upper limit on the PBH abundance from each analysis and do not statistically combine the three samples.

We derive conservative upper limits on the present-day PBH abundance using a background-agnostic treatment as in Ref.~\cite{Chianese:2024rsn}. The published segmented flux measurements and event counts are treated as upper bounds on an additional PBH contribution. Therefore, a flux or event-count prediction contributes only when it exceeds the corresponding observed value. Predictions below the observed value give no penalty, since the remaining flux or events may originate from non PBH sources that are not modeled explicitly.

For each bin \(i\), let \(d_i\) denote the measured input and let \(m_i(\boldsymbol{\theta})\) denote the corresponding PBH prediction. The model parameters are \(\boldsymbol{\theta}_{\rm mono} = (M_0, \fpbh, k)\) for a monochromatic PBH population and \(\boldsymbol{\theta}_{\rm LN} = (M_c, \fpbh, \sigma, k)\) for a log-normal PBH population. For HESE and MESE, \(m_i(\boldsymbol{\theta})\) is a binned flux measurement, whereas for EHE it is an expected event count. When deriving an upper limit at each mass point, \(M_0\) or \(M_c\), \(k\), and where applicable, \(\sigma\), are held fixed while \(\fpbh\) is varied. The PBH abundance is therefore the sole parameter of interest in this analysis. We define the background-agnostic likelihood function as
\begin{equation}
\mathcal{L}(\boldsymbol{\theta})
=
\prod_i
\begin{cases}
\dfrac{\mathcal{P}\!\left(d_i\mid m_i(\boldsymbol{\theta})\right)}
      {\mathcal{P}\!\left(d_i\mid d_i\right)},
& m_i(\boldsymbol{\theta})>d_i, \\
1,
& m_i(\boldsymbol{\theta})\leq d_i ,
\end{cases}
\label{eq:lambda_total}
\end{equation}
where \(\mathcal{P}\) is the probability distribution for data \(d_i\) with expected mean \(m_i(\boldsymbol{\theta})\), and the index \(i\) runs over the number of data points.
For the HESE and MESE segmented flux measurements, the probability distribution \(\mathcal{P}\) is taken to be Gaussian. We thus adopt the test statistic
\begin{equation}
\rm TS(\boldsymbol{\theta})=-2\ln\mathcal
{L}(\boldsymbol{\theta}).
\label{eq:likelihood_ratio_stat}
\end{equation}
The PBH prediction in bin \(i\) is computed as the logarithmic-bin average
\begin{align}
m_i(\boldsymbol{\theta})
=
\frac{1}{\Delta \log_{10} E_i}
\int_{\log_{10} E_i^{\rm min}}^{\log_{10} E_i^{\rm max}}
d(\log_{10}E_\nu)\,
E_\nu^2
\frac{d^2\Phi_\nu(E_\nu,\boldsymbol{\theta})}{dE_\nu\,d\Omega},
\end{align}
where
\(\Delta \log_{10} E_i =
\log_{10}\left(\frac{E_i^{\rm max}}{E_i^{\rm min}}\right)\).
Since the one-sided test statistic only accounts for an additional positive PBH contribution, we use the quoted upper uncertainty \(\sigma_i^+\) in each bin. We further neglect any correlations between the energy bins. The resulting contribution to the test statistic is therefore
\begin{equation}
\rm TS(\boldsymbol{\theta}) =
\begin{cases}
\left[\dfrac{m_i(\boldsymbol{\theta})-d_i}{\sigma_i^+}\right]^2,
& m_i(\boldsymbol{\theta})>d_i,\\
0, & m_i(\boldsymbol{\theta})\le d_i .
\end{cases}
\label{eq:gaussian_ts}
\end{equation}

For IceCube EHE and for the future event-count analyses, we instead write the predicted number of events in detector \(d\) and energy bin \(j\) as
\begin{equation}
\mu_{dj}(\boldsymbol{\theta})
=
\sum_{\alpha=e,\mu,\tau}
\int_{E_j^{\min}}^{E_j^{\max}}
\dd E_\nu\,
\frac{\dd^2\Phi_{\nu_\alpha}(E_\nu,\boldsymbol{\theta})}
     {\dd E_\nu\,\dd\Omega}\,
\mathcal{E}_{d,\alpha}(E_\nu),
\label{eq:general_event_count}
\end{equation}
where \(\mathcal{E}_{d,\alpha}(E_\nu)\) is the exposure for flavor \(\alpha\). Here, we write the PBH diffuse neutrino flux as all-flavor, \(\Phi_\nu^{\rm all}=\sum_\alpha \Phi_{\nu_\alpha}\), and assume a \(1:1:1\) flavor ratio at Earth, so that \(\Phi_{\nu_\alpha}=\Phi_\nu^{\rm all}/3\). For the IceCube EHE analysis, the published exposure is given as a three-flavor-summed exposure
\begin{equation}
\mathcal{E}_{\rm EHE, 3{\rm f}}(E)
=
\sum_{\alpha=e,\mu,\tau}
\mathcal{E}_{{\rm EHE},\alpha}(E_\nu).
\label{eq:ehe_3flavor_exposure}
\end{equation}
Accordingly, when evaluating Eq.
\eqref{eq:general_event_count} with the three-flavor-summed EHE exposure, we use the per-flavor flux \(\Phi_\nu^{\rm all}/3\). Furthermore, because of the lower-statistics, we take the probability distribution \(\mathcal{P}\) to be Poissonian in this case. We apply the same one-sided construction with \(m_i(\boldsymbol{\theta})=\mu_{\rm EHE}(\boldsymbol{\theta})\), and \(d_i=n_{\rm obs}=1\), yielding the test statistic
\begin{equation}
\rm{TS}_{\rm EHE}(\boldsymbol{\theta}) =
\begin{cases}
2\left[
\mu_{\rm EHE}(\boldsymbol{\theta})-1
-\ln(\mu_{\rm EHE}(\boldsymbol{\theta}))
\right],
& \mu_{\rm EHE}>1,\\
0, & \mu_{\rm EHE}\le 1.
\end{cases}
\label{eq:ehe_poisson_ts}
\end{equation}
We approximate the distribution of these one-sided test statistics using the asymptotic half-\(\chi_1^2\) result of Ref. \cite{Cowan:2010js}. The corresponding critical value for a one-sided \(90\%\) confidence limit with one degree of freedom is \(\Delta \chi^2 = 1.64\).

\subsection{Future detector forecasts}
\label{sec:future_projections}

Several proposals for future neutrino detectors aim to extend searches into the ultra-high-energy range, such as RNO-G \cite{RNO-G:2020rmc}, POEMMA \cite{POEMMA:2020ykm}, Trinity \cite{Brown:2021ane}, IceCube-Gen2 radio \cite{IceCube:2019pna}, and GRAND200k \cite{GRAND:2018iaj}. Many of these experiments aim to detect secondary radio signals produced by high-energy neutrinos. RNO-G and IceCube-Gen2 radio are designed detect Askaryan radio pulses from neutrino-induced particle cascades in ice. GRAND200k is designed to detect radio emission from air showers initiated by tau leptons produced in Earth-skimming \(\nu_\tau\) interactions, while POEMMA and Trinity search for Cherenkov light from tau-induced air showers. In this work, we focus on IceCube-Gen2 radio and GRAND200k as representative future radio-based benchmarks for ultra-high-energy neutrino searches. Representative projected 90\% CL differential all-flavor sensitivities for IceCube-Gen2 radio and GRAND200k are shown for reference in Fig.~\ref{fig:benchmark-flux-data}.

For the IceCube-Gen2 radio forecast, we use the radio-only flavor-dependent effective volumes of Ref.~\cite{IceCube-Gen2:2023pyn}. These effective volumes are direction averaged and include attenuation of the neutrino flux in the earth. Since the radio-only curves include both charged-current and neutral-current interactions, the effective area is defined as
\begin{equation}
A_{\rm eff, \alpha}^{\rm radio}(E_\nu)
=
4\pi V_{{\rm eff},\alpha}^{\rm radio}(E_\nu)\,
n_N\!\left[
\sigma_{\nu N}^{\rm CC}(E_\nu)
+\sigma_{\nu N}^{\rm NC}(E_\nu)
\right],
\label{eq:veff_to_aeff}
\end{equation}
where \(\alpha=e,\mu,\tau\), \(\sigma_{\nu N}^{\rm CC}\) and \(\sigma_{\nu N}^{\rm NC}\) are the neutrino-nucleon charged-current
and neutral-current cross sections, respectively, and \(n_N\) is the
number density of nucleons in ice. We use the central neutrino-nucleon cross-section parametrization of Ref.~\cite{Connolly:2011vc} and take
\(n_N=\rho_{\rm ice}N_A\), with \(\rho_{\rm ice}=0.92\,{\rm g\,cm^{-3}}\). The corresponding flavor-dependent IceCube-Gen2 radio exposure is
\begin{equation}
\mathcal{E}_{{\rm Gen2},\alpha}(E_\nu)
=
T_{\rm Gen2}
A_{\rm eff, \alpha}^{\rm radio}(E_\nu).
\label{eq:gen2_flavor_exposure}
\end{equation}

For GRAND200k, we use the published GRAND10k HotSpot1 exposure in Ref.~\cite{GRAND:2018iaj}. We use the aggressive-threshold HotSpot1 exposure and assume that it scales linearly with detector size and live-time. The scaled GRAND200k \(\nu_\tau\) exposure is then 
\begin{equation}
\mathcal{E}_{\rm GRAND200k, \tau}(E_\nu)
= 20 \times
\left(\frac{T_{\rm GRAND10k}}{1\,{\rm yr}}\right)
\,
\mathcal{E}_{\rm HS1, \tau}^{1{\rm yr}}(E_\nu),
\label{eq:grand_tau_exposure}
\end{equation}
where \(\mathcal{E}_{\rm HS1, \tau}^{1{\rm yr}}(E_\nu)\) is the one-year GRAND10k HotSpot1 exposure. The factor of \(20\) scales the GRAND10k exposure to GRAND200k, while \(T_{\rm GRAND}/1\,{\rm yr}\) accounts for the assumed live-time. Since GRAND200k is primarily sensitive to Earth-skimming \(\nu_\tau\) events, the
GRAND200k event count is evaluated using the \(\nu_\tau\) component of the all-flavor PBH flux.

Future event counts are then evaluated using Eq. \eqref{eq:general_event_count}, with
the exposures for both GRAND200k and IceCube-Gen2 radio. Unless otherwise stated, we take
\(T_{\rm Gen2}=T_{\rm GRAND}=10\,\mathrm{yr}\). To compute the projected upper limits we follow Ref.~\cite{Chianese:2024rsn}, and adopt a background-free zero-count scenario. The projected 90\% CL upper limit for a detector is then defined by 
\begin{equation}
\sum_j \mu_{dj}(\boldsymbol{\theta})=2.44,
\label{eq:fc_boundary}
\end{equation}
in accordance with the Feldman-Cousins limit for \(n_{\rm obs}=0\) with zero background \cite{Feldman:1997qc}.

\subsection{Bayesian forecast method}
\label{sec:bayesian_methods}

The analyses above use a frequentist approach to derive upper limits on the present-day PBH abundance from current data and from projected future null observations. These limits determine which PBH populations are disfavored when no PBH-induced neutrino signal has been established.

We next consider a complementary forecast in which a PBH-induced neutrino signal is assumed to be present. The aim is to reconstruct the properties of the PBH population and determine whether the data favor a monochromatic or a log-normal initial mass function. The total event count alone is generally insufficient because variations in the abundance, mass, memory burden parameter and mass-function width can partially compensate one another. The event spectra in IceCube-Gen2 radio and GRAND200k therefore provide the main information for parameter reconstruction and mass-function discrimination.

Both frequentist and Bayesian methods could in principle be used for the upper limits or signal inference. Here, we choose to adopt a Bayesian framework for the simulated-signal analysis because it provides a unified treatment of multidimensional parameter reconstruction and model comparison. Posterior distributions are used to summarize the constraints and degeneracies among the PBH parameters, while Bayesian evidences are used to compare the monochromatic and log-normal hypotheses. Furthermore, in contrast to the frequentist analysis above, the Bayesian credible regions are obtained directly from the posterior distribution and do not require an assumed asymptotic distribution for a test statistic making it suitable for smaller signal data sets. 

The simulated event samples are generated from the benchmark parameter points in Table \ref{tab:benchmarks}. We define separate benchmark parameter points for the monochromatic and log-normal populations so that each mass-function family can be tested both as the data-generating model and as a fitted hypothesis. We consider \(k_{\rm true}=2\) and \(k_{\rm true}=4\) as benchmark cases of moderate and stronger memory burden suppression. These choices lead to distinct evaporation histories while allowing surviving PBHs to emit neutrinos in the energy range relevant for IceCube-Gen2 radio and GRAND200k. For the log-normal benchmark points, we set \(\sigma_{\rm true}=1\), representing a broad mass distribution that is clearly distinct from the monochromatic case. 

For each benchmark point, the mass parameter is selected in a region where the future detectors are sensitive. The PBH abundance is then chosen to give 30 expected events in the combined IceCube-Gen2 radio and GRAND200k exposure, while remaining compatible with the current IceCube upper limits. We choose this event yield because it provides adequate spectral information for meaningful parameter reconstruction and mass-function discrimination. We emphasize that these benchmarks are not predictions of the event rate expected in the future experiments, but rather show what could be concluded from a signal of this size.

The simulated event samples are generated under a signal-only approximation. The event counts are assumed to arise from the PBH signal alone, with no additional background components. The resulting forecasts therefore describe the ideal information contained in the PBH signal. Further consideration of astrophysical and detector backgrounds would weaken parameter reconstruction and mass-function discrimination. For a benchmark parameter point \(\boldsymbol{\theta}_b\), the simulated counts are drawn as
\begin{equation} 
n_{dj} \sim {\rm Poisson}\!\left(\mu_{dj}(\boldsymbol{\theta}_{\rm b})\right), \label{eq:bayes_simulated_counts}
\end{equation} 
and the resulting dataset is denoted by \(D=\{n_{dj}\}\).

For the Bayesian model comparison, we generate 50 Poisson realizations of every benchmark and summarize the resulting distribution of Bayes factors. Parameter reconstruction is instead performed once for each benchmark using the expected-count dataset, \(n_{dj}=\mu_{dj}(\boldsymbol{\theta}_{\rm b})\), so the corner plots provide a representative fluctuation free forecast rather than the result of a particular Poisson realization.

For the analysis, we start from Bayes' theorem \cite{Trotta:2008qt}, 
\begin{equation} 
P(\boldsymbol{\boldsymbol{\theta}} \lvert D, \mathcal{M}) = \frac{\mathcal{L}(D \lvert \boldsymbol{\boldsymbol{\theta}}, \mathcal{M})\pi(\boldsymbol{\boldsymbol{\theta}} \lvert \mathcal{M})}{Z_{\mathcal{M}}}
\label{eq:bayes_theorem} 
\end{equation}
where \(P(\boldsymbol{\boldsymbol{\theta}} \lvert D, \mathcal{M})\) is the posterior probability distribution of the model parameters \(\boldsymbol{\boldsymbol{\theta}}\), \(D\) denotes the data, and \(\mathcal{M}\) denotes the model under consideration. Furthermore, \(Z_{\mathcal{M}}\) is the evidence, \(\mathcal{L}(D\lvert\boldsymbol{\boldsymbol{\theta}},\mathcal{M})\) is the likelihood, and \(\pi(\boldsymbol{\boldsymbol{\theta}}\lvert\mathcal{M})\) is the prior probability distribution.

For the Bayesian likelihood, each detector response is divided into five logarithmically spaced bins over its tabulated energy range. The bins are defined in true neutrino energy, and detector energy resolution smearing is not included. For a fitted mass-function model \(\mathcal{M}\), the likelihood is taken to be Poissonian, with the detector and energy bins treated as statistically independent, giving
\begin{equation} 
\ln \mathcal{L}(D|\boldsymbol{\theta},\mathcal{M}) = \sum_{dj}[n_{dj}\ln{\mu_{dj}(\boldsymbol{\theta})} - \mu_{dj}(\boldsymbol{\theta}) - \ln \Gamma(n_{dj}+1)].
\label{eq:poisson_likelihood} 
\end{equation}
For individual-detector forecasts, the product over \(d\) contains either IceCube-Gen2 radio or GRAND200k, whereas for the combined forecast, it contains both detectors. Furthermore, the Bayesian evidence for a model \(\mathcal{M}\) is given by the marginalized likelihood
\begin{equation} 
Z_{\mathcal{M}} = P(D|\mathcal{M}) = \int\dd\boldsymbol{\theta}\, \mathcal{L}(D|\boldsymbol{\theta},\mathcal{M}) \pi(\boldsymbol{\theta}|\mathcal{M}), 
\label{eq:evidence} 
\end{equation} 

Posterior samples and Bayesian evidences are numerically estimated with the nested sampling algorithm \textsc{UltraNest} \cite{Skilling:2006gxv,Buchner:2021cql}. All reported estimates satisfy the acceptance requirements summarized in Appendix \ref{app:priors}.

The mass-function discrimination analysis is performed by fitting the same simulated dataset with both the monochromatic and log-normal models, independently of which model generated the dataset. The model preference is quantified by the Bayes factor
\begin{equation} 
B_{\rm LN,mono} = \frac{Z_{\rm LN}}{Z_{\rm mono}}, \label{eq:bayes_factor} 
\end{equation}
or equivalently by its logarithm, 
\begin{equation}
\ln B_{\rm LN,mono} = \ln Z_{\rm LN}-\ln Z_{\rm mono}. 
\label{eq:delta_ln_z} 
\end{equation}
With this convention, \(\ln B_{\rm LN,mono}>0\) favors the log-normal model, while \(\ln B_{\rm LN,mono}<0\) favors the monochromatic model. We use the Bayes-factor scale of Ref.~\cite{Kass:1995loi} as a qualitative guide when interpreting the Bayesian model-comparison results.
We therefore regard \(|\ln B_{\rm LN,mono}|<1\) as inconclusive, \(1<|\ln B_{\rm LN,mono}|<3\) as positive evidence, \(3<|\ln B_{\rm LN,mono}|<5\) as strong evidence, and \(|\ln B_{\rm LN,mono}|>5\) as very strong evidence.

Parameter reconstruction is treated separately from mass-function discrimination. In this case, the mass-function family is assumed to be known, while its parameters are inferred. Therefore, a simulated dataset generated from a monochromatic benchmark parameter point is fitted only with the monochromatic model, while a simulated dataset generated from a log-normal benchmark parameter point is fitted only with the log-normal model. The output is the marginalized posterior distribution of the PBH parameters within the assumed mass distribution family. 

The priors used for Bayesian model comparison and parameter reconstruction are summarized in Tables~\ref{tab:priors_model_comparison} and
\ref{tab:priors_reconstruction}. These priors should be understood as tied to the local forecast detector-sensitive mass windows, rather than encapsulating the complete PBH parameter space. For the model comparison, parameters shared by the two hypotheses are assigned identical prior ranges. 

\begin{table}[t]
\centering
\caption{Benchmark PBH parameter points used to generate the simulated event samples in the Bayesian forecast. For each benchmark point, \(\fpbh^{\rm true}\) is chosen to give 30 events in the combined IceCube-Gen2 radio and
GRAND200k exposure.}
\label{tab:benchmarks}
\begin{ruledtabular}
\begin{tabular}{@{}lcccc@{}}
Mass function & \(k_{\rm true}\) & Mass parameter & \(\sigma_{\rm true}\) & \(\fpbh^{\rm true}\) \\
\hline
Monochromatic
& \(2\)
& \(M_0^{\rm true}=5.62\times10^4\g\)
& - & \(7.54\times10^{-5}\) \\
Log-normal
& \(2\)
& \(M_c^{\rm true}=3.16\times10^5\g\)
& \(1.0\) & \(7.45\times10^{-8}\) \\
Monochromatic
& \(4\)
& \(M_0^{\rm true}=1.88\times10^1\g\)
& - & \(3.73\times10^{-4}\) \\
Log-normal
& \(4\)
& \(M_c^{\rm true}=1.88\times10^2\g\)
& \(1.0\) & \(3.85\times10^{-7}\) \\
\end{tabular}
\end{ruledtabular}
\end{table}

\section{Results}
\label{sec:results}

\subsection{Current and projected PBH abundance limits}
\label{sec:abundance_limits}

\begin{figure*}[t]
    \centering
    \includegraphics[width=\textwidth]{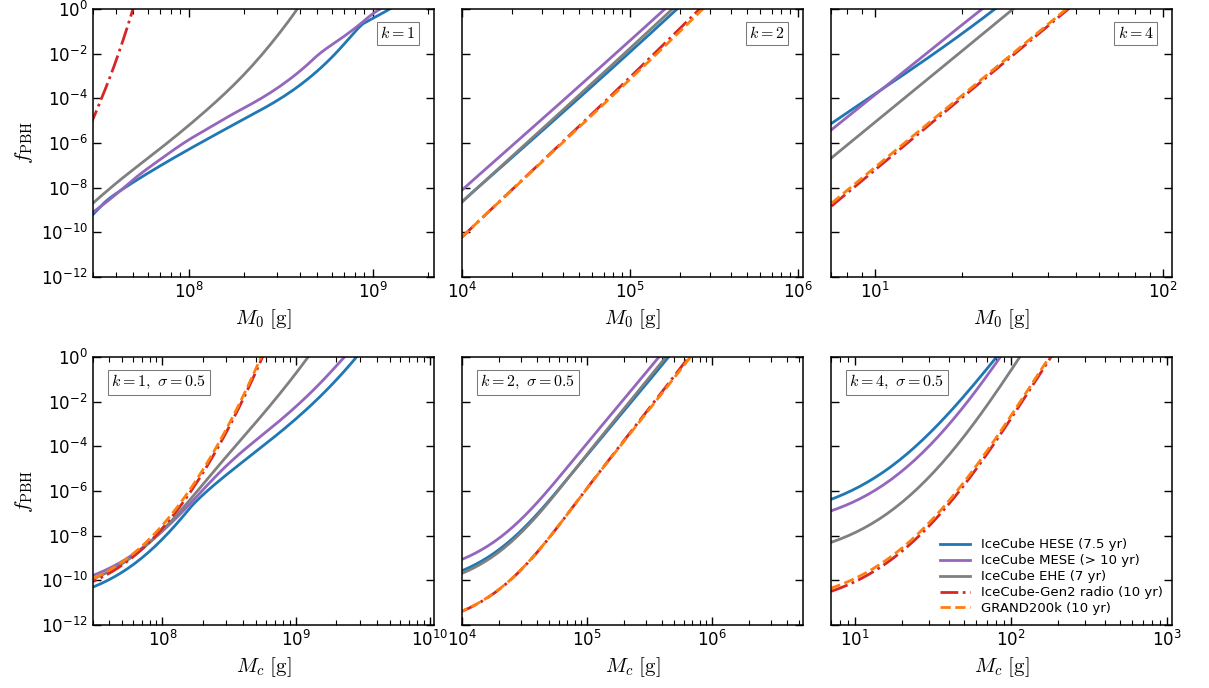}
    \caption{90\% CL upper limits on the present-day PBH abundance \(\fpbh\) from the current IceCube HESE, MESE, and EHE datasets, together with the projected upper limits for IceCube-Gen2 radio and GRAND200k assuming 10 years of exposure. The top row shows the results for a monochromatic initial mass function, while the bottom row shows the results for a log-normal initial mass function with \(\sigma = 0.5\). The columns correspond to \(k=1\), \(2\), and \(4\). Abundances above each curve are excluded.}
    \label{fig:abundance-limits}
\end{figure*}

\begin{figure*}[t]
    \centering
    \includegraphics[width=\textwidth]{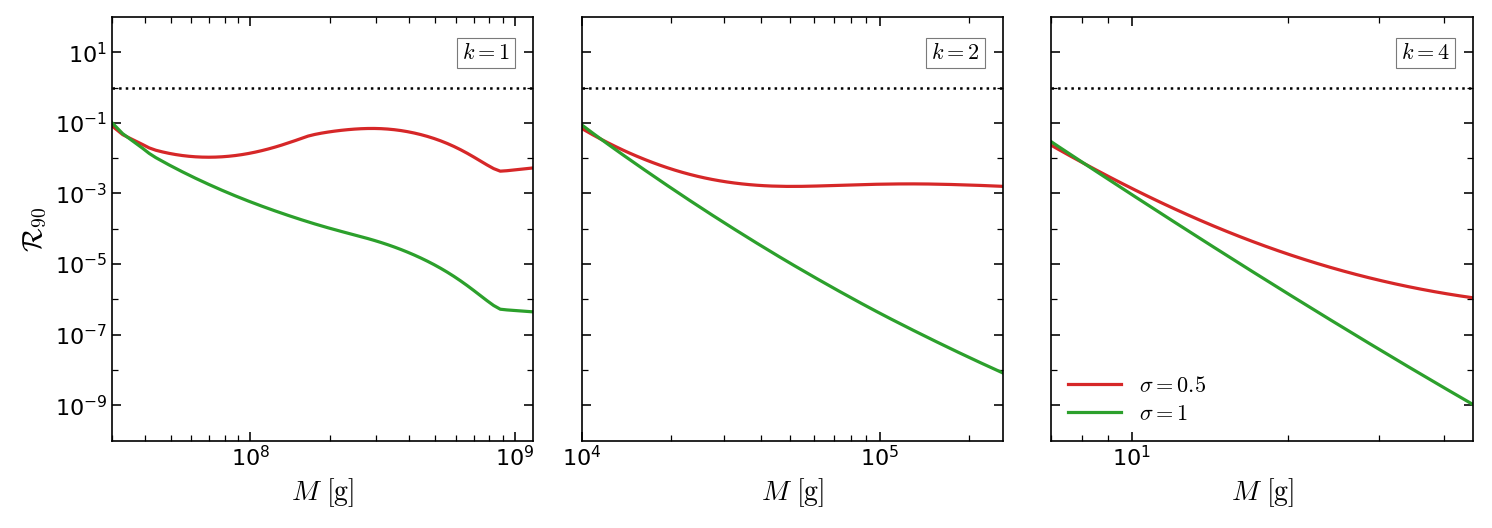}
    \caption{Ratio \(\mathcal{R}_{90}\) of the strongest \(90\%\) CL upper limits on \(\fpbh\) obtained for log-normal and monochromatic initial mass functions, evaluated at equal characteristic mass, \(M\equiv M_0 = M_c\). The panels correspond to \(k=1\), \(2\), and \(4\), while the curves show the results for \(\sigma= 0.5\) and \(1\). Values below unity indicate that the log-normal initial mass function gives a stronger constraint than the corresponding monochromatic mass function. The dotted horizontal line marks \(\mathcal{R}_{90}=1\).}
    \label{fig:limit-ratio}
\end{figure*}

We first present the frequentist upper limits on the present-day PBH abundance obtained from the current IceCube datasets, together with the projected upper limits for IceCube-Gen2 radio and GRAND200k assuming 10 years of exposure. The results are shown in Fig.~\ref{fig:abundance-limits} which compares the \(90\%\) CL abundance
limits for the different datasets.

For all values of \(k\), the limits generally weaken toward larger characteristic masses. This is expected because more massive PBHs have lower Hawking temperatures and thus smaller neutrino emission rates, as seen in Eqs. \eqref{eq:HawkingTemperature} and \eqref{eq:hawking_spectrum}. A larger PBH abundance is therefore required to produce the same neutrino flux. Increasing \(k\) shifts the constrained mass range toward lower masses because the stronger suppression of late-time evaporation allows lighter PBHs to survive until the present epoch.

In the case of \(k=1\), the strongest constraints are obtained from the current IceCube datasets both for the monochromatic and log-normal results. The HESE sample provides the leading limit over most of the mass range shown, while the MESE and EHE constraints are generally weaker. The projected IceCube-Gen2 radio and GRAND200k upper limits do not improve on the current IceCube limits in this case. This is because the surviving PBHs are relatively massive, so their neutrino emission falls in the energy range covered by the current IceCube data.

The situation however changes for \(k=2\) and \(k=4\). The lower surviving PBH masses correspond to higher Hawking temperatures, shifting the emitted neutrino spectrum toward the energy range probed by the radio detectors. As a result, the projected IceCube-Gen2 radio and GRAND200k upper limits are considerably stronger than the current IceCube constraints over the relevant mass range. Among the current IceCube datasets, the HESE and EHE limits are comparable for \(k=2\), while the EHE sample gives the strongest current constraint for \(k=4\).

Generally, the log-normal results show the same dependence on \(k\) and the same change in the constraining power of the datasets. At a fixed characteristic mass, the log-normal distribution generally gives stronger limits than the corresponding monochromatic distribution. It also extends the constraints over a broader range of characteristic masses. For the current IceCube datasets, the strongest upper limit is obtained for \(k=1\) near a characteristic mass of \(3\times 10^7\,{\rm g}\) and are approximately \(\fpbh\simeq 10^{-9}\) for the monochromatic population and \(\fpbh\simeq9\times10^{-11}\) for the log-normal population with \(\sigma=0.5\). By contrast, the strongest projected upper limits occur for \(k=2\) near a characteristic mass of
\(10^4\,{\rm g}\) and are approximately \(\fpbh\simeq 10^{-10}\) for the monochromatic population and \(\fpbh\simeq6\times10^{-12}\) for the log-normal population with \(\sigma=0.5\). The difference between the monochromatic and log-normal upper limits is examined more directly in Sec. \ref{sec:width_impact}.

\subsection{Impact of the extended mass function}
\label{sec:width_impact}

\begin{figure*}[t]
    \centering
    \includegraphics[width=\textwidth]{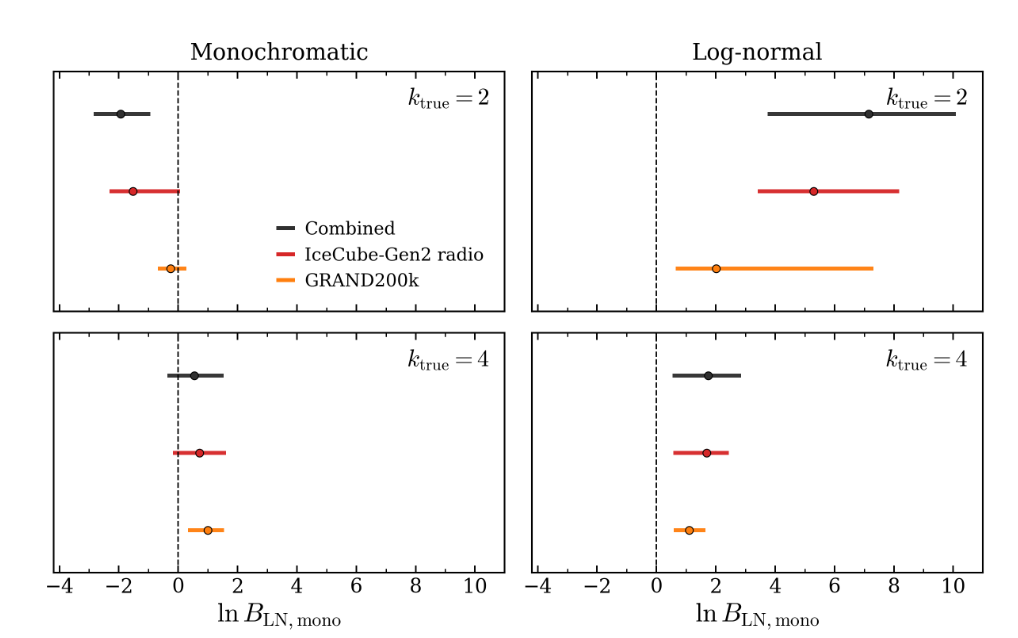}
    \caption{Bayesian model comparison of the monochromatic and log-normal mass-function hypotheses for simulated IceCube-Gen2 radio and GRAND200k event samples. The left and right columns correspond to the signal being generated under the monochromatic and log-normal mass-function hypotheses, respectively, while the top and bottom rows correspond to \(k_{\rm true}=2\) and \(4\). The points show the median log Bayes factor, \(\ln B_{\rm LN,mono}=\ln Z_{\rm LN}-\ln Z_{\rm mono}\), over 50 Poisson realizations and the horizontal bars show the corresponding \(68\%\) intervals. Results are shown for the combined dataset and for each detector separately. Positive values favor the log-normal model, negative values favor the monochromatic model, and the vertical dashed line marks equal evidence.}
    \label{fig:mass-function-comparison}
\end{figure*}

\begin{figure*}[t]
    \centering
    \includegraphics[width=\textwidth]{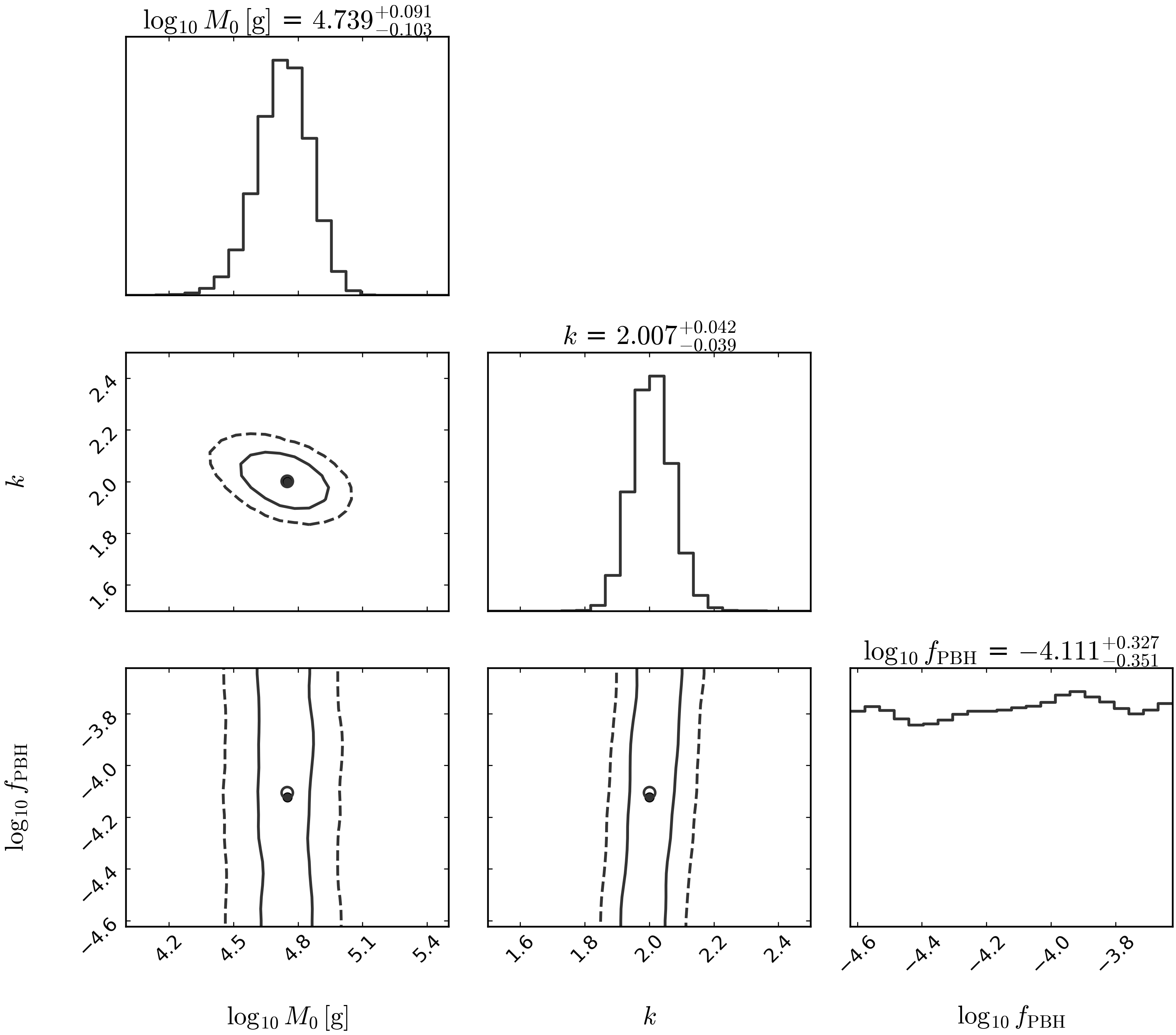}
    \caption{Posterior reconstruction for the monochromatic benchmark with \(k_{\rm true}=2\), \(M_0^{\rm true}=5.62\times10^4\,{\rm g}\), and \(\fpbh^{\rm true}=7.54\times10^{-5}\), using the combined IceCube-Gen2 radio and GRAND200k expected-count dataset containing 30 signal events. The diagonal panels show the marginalized one-dimensional posteriors, and the off-diagonal panels show the joint posterior distributions. Solid and dashed contours enclose \(68\%\) and \(95\%\) posterior probability, respectively. Filled circles indicate the benchmark parameters and open circles indicate the maximum-likelihood point.}
    \label{fig:mono-k2-reconstruction}
\end{figure*}

\begin{figure*}[t]
    \centering
    \includegraphics[width=\textwidth]{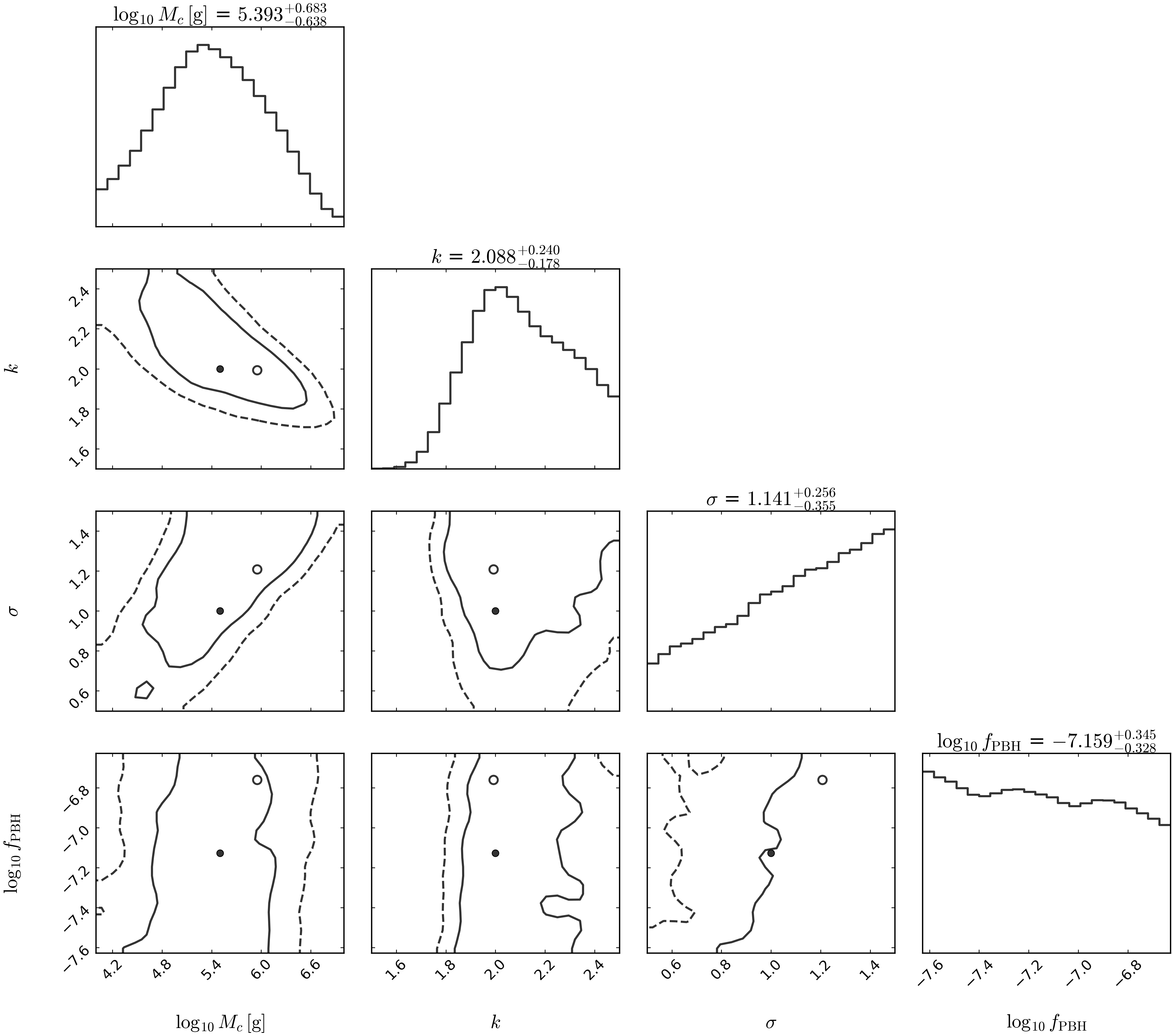}
    \caption{Posterior reconstruction for the log-normal benchmark with \(k_{\rm true}=2\), \(M_c^{\rm true}=3.16\times10^5\,{\rm g}\),
    \(\sigma_{\rm true}=1\), and \(\fpbh^{\rm true}=7.45\times10^{-8}\), using the combined IceCube-Gen2 radio and GRAND200k expected-count dataset containing 30 signal events. The plotting conventions are the same as in Fig. \ref{fig:mono-k2-reconstruction}, with the addition of the log-normal width parameter \(\sigma\).}
    \label{fig:lognormal-k2-reconstruction}
\end{figure*}

To quantify the effect of the extended mass function on the abundance limits, we define the ratio
\begin{equation}
\mathcal{R}_{90}(M)
= \frac{\displaystyle \min_d f_{\rm PBH,LN}^{90\%,d}(M_c=M) }{ \displaystyle \min_d f_{\rm PBH,mono}^{90\%,d}(M_0=M)},
\label{eq:limit_ratio}
\end{equation}
where \(d\) runs over the current IceCube datasets and the projected IceCube-Gen2 radio and GRAND200k results. The comparison is performed at equal characteristic mass, \(M\equiv M_0=M_c\), with \(k\) and \(\sigma\) specified by the corresponding panel and curve, respectively. The numerator and denominator therefore represent the strongest upper limit obtained at each mass for the log-normal and monochromatic initial mass functions. Values of \(\mathcal{R}_{90}<1\) indicate that the log-normal initial mass function gives a stronger upper limit on \(\fpbh\) than the corresponding monochromatic mass function. Fig. \ref{fig:limit-ratio} shows that the log-normal mass function gives stronger limits throughout the mass ranges considered in this work. For \(\sigma=0.5\) and \(\sigma=1\), the ratio generally decreases with increasing mass, particularly for \(k=2\) and \(k=4\). Near the upper ends of the displayed mass ranges, \(\mathcal{R}_{90}\) reaches values of approximately \(10^{-2}\)--\(10^{-6}\) for \(\sigma=0.5\) and \(10^{-6}\)--\(10^{-9}\) for \(\sigma=1\), depending on \(k\). The corresponding log-normal limits can therefore be stronger than the monochromatic limits by several orders of magnitude. For \(k=1\), the ratio has a less monotonic mass dependence but remains below unity for all widths. The strengthening of the limits is caused by the low-mass tail of the log-normal distribution. PBHs in this tail have higher Hawking temperatures and emit neutrinos more efficiently than the PBHs at the median mass. Their contribution can therefore dominate the predicted signal, particularly for broader mass functions.

\subsection{Bayesian mass-function discrimination}
\label{sec:bayesian_comparison}

We next examine whether the simulated event distributions can distinguish between the monochromatic and log-normal mass-function hypotheses. Fig.~\ref{fig:mass-function-comparison} summarizes the resulting log Bayes factors for the four benchmark signals.

For \(k_{\rm true}=2\), the result depends strongly on the underlying mass function. For the monochromatic benchmark, all three median log Bayes factors favor the correct model in sign. The median is approximately \(-0.26\) for GRAND200k, \(-1.53\) for IceCube-Gen2 radio, and \(-1.94\) for the combined dataset. The GRAND200k result remains inconclusive, while the IceCube-Gen2 radio and combined medians provide positive support for the monochromatic hypothesis. The \(68\%\) interval for the combined dataset remains negative, whereas those of the individual detectors extend across or are close to equal evidence. For the log-normal benchmark, all three dataset choices favor the correct model. The median is approximately \(2.01\) for GRAND200k, \(5.30\) for IceCube-Gen2 radio, and \(7.15\) for the combined dataset. The GRAND200k data typically provides positive evidence, while the IceCube-Gen2 radio and combined dataset provide very strong evidence in this scenario. The broad and asymmetric \(68\%\) intervals, particularly for GRAND200k, show that the strength of the preference varies substantially between Poisson realizations. We also remark that these results significantly depends on the choice of prior. As examined in Appendix~\ref{app:prior_sensitivity}, particularly extending the mass range prior can reduce the combined median to \(\ln B_{\rm LN,mono}=1.43^{+0.63}_{-0.49}\), while preserving its positive sign. 

For \(k_{\rm true}=4\), all three dataset choices give positive median log Bayes factors for both underlying mass functions. For the monochromatic benchmark, the medians are approximately \(1.00\) for GRAND200k, \(0.72\) for IceCube-Gen2 radio, and \(0.54\) for the combined dataset. These values incorrectly favor the log-normal hypothesis in sign, although they remain in the inconclusive-evidence category. For the log-normal benchmark, the corresponding medians are approximately \(1.10\), \(1.69\), and \(1.75\), respectively, correctly favoring the log-normal hypothesis with positive evidence. The underlying mass function is therefore not reliably identified for \(k_{\rm true}=4\), indicating a stronger spectral degeneracy between the two hypotheses.

Overall, these results show that the broad log-normal mass distribution can generally be correctly distinguished from the monochromatic distribution, although a conservative estimate from 30 events provides only inconclusive to slightly positive evidence. On the contrary, when the true distribution is taken to be monochromatic the evidence can yield the wrong conclusion. The inherent degeneracy lies in that a monochromatic PBH population with a strong memory burden suppression of \(k=4\) closely mimics the signature of the dominant high energy tail signal of an extended mass distribution.

\subsection{Bayesian parameter reconstruction}
\label{sec:parameter_reconstruction}

We next examine parameter reconstruction using the combined IceCube-Gen2 radio and GRAND200k expected-count dataset containing
30 signal events. Figs.~\ref{fig:mono-k2-reconstruction} and \ref{fig:lognormal-k2-reconstruction} show the marginalized posterior distributions for the \(k_{\rm true}=2\) monochromatic and log-normal benchmarks, respectively.

For the monochromatic benchmark, the mass and memory-burden parameters are reconstructed accurately. The marginalized \(68\%\) credible intervals are
\begin{equation}
\log_{10}\!\left(\frac{M_0}{\mathrm{g}}\right)
=
4.739^{+0.091}_{-0.103},
\qquad
k=2.007^{+0.042}_{-0.039},
\end{equation}
consistent with the benchmark values \(\log_{10}(M_0^{\rm true}/{\rm g})=4.750\) and \(k_{\rm true}=2\). Furthermore, we have a mild anticorrelation between \(M_0\) and \(k\), indicating that variation in each of these parameters is interchangeable in the event spectrum. Nevertheless, both parameters are reconstructed well.

The log-normal reconstruction is substantially less constraining because the additional width parameter introduces further spectral degeneracies. The marginalized \(68\%\) credible intervals are
\begin{equation}
\begin{aligned}
    \log_{10}\!\left(\frac{M_c}{\mathrm{g}}\right) &=
    5.393^{+0.683}_{-0.638}, \\
    k &=  2.088^{+0.240}_{-0.178}, \\
    \sigma &= 1.141^{+0.256}_{-0.355}.
\end{aligned}
\label{eq:lognorm-1sigma}
\end{equation}
The marginalized posterior for \(k\) is asymmetric, with a broader tail toward larger values. The posterior for \(\sigma\) also retains substantial support toward the upper edge of the adopted prior, \(\sigma = 1.5\), so its upper uncertainty should be interpreted within the adopted prior rather than as a closed upper bound. All three benchmark values nevertheless lie within the corresponding credible regions. 

The elongated contours show that an increase in \(M_c\) can be partially compensated by changes in \(k\) and \(\sigma\). In particular, the central mass and width are positively correlated, while \(M_c\) and \(k\) are anticorrelated. Consequently, the characteristic mass is determined only to approximately \(0.7\) dex, and the width remains comparatively weakly constrained.

In both reconstructions, the marginalized posterior for \(\log_{10}\fpbh\) remains broad across the adopted local prior. The abundance is therefore not precisely reconstructed once the other model parameters are marginalized over, and its quoted median should not be interpreted as a precise measurement. Overall, \(M_0\) and \(k\) are constrained precisely for a monochromatic population, whereas the additional freedom of the log-normal model produces substantially broader and more correlated parameter constraints.

For comparison, the corresponding posterior reconstructions for \(k_{\rm true}=4\) are presented in Appendix \ref{app:k4-reconstruction}. These reconstructions are considerably weaker than those obtained for \(k_{\rm true}=2\). In the monochromatic case, \(M_0\) and \(k\) follow a broad anticorrelated band while in the log-normal case, this degeneracy is joined by correlations with \(\sigma\). At larger \(k\), the mass and memory burden parameter have more strongly coupled effects on the PBH evolution. Changes in these parameters can therefore compensate one another and produce similar event count spectra. The likelihood consequently identifies a broad range of parameter combinations, although the benchmark values remain within the corresponding credible regions.

\section{Discussion and Conclusions}
\label{sec:discussion}

In this work, we studied the high- and ultra-high-energy neutrino signatures of memory-burdened PBHs with monochromatic and log-normal initial mass functions. We derived \(90\%\) CL upper limits on the present-day PBH abundance from the current IceCube HESE, MESE, and EHE datasets alongside the projected 10 year sensitivity of IceCube-Gen2 radio and GRAND200k. Current IceCube data provide the leading limits for \(k=1\), whereas the projected radio detectors become substantially more sensitive for \(k=2\) and \(k=4\). This change occurs because stronger memory burden suppression allows lighter and hotter PBHs to survive until today, shifting their emission toward the energies probed by the radio detectors. Across the parameter ranges considered, the strongest projected upper limits occur for \(k=2\) near a characteristic mass of \(10^4\,{\rm g}\) and are approximately \(\fpbh\simeq 10^{-10}\) for the monochromatic population and \(\fpbh\simeq6\times10^{-12}\) for the log-normal population with \(\sigma=0.5\). However, for the current IceCube constraints, the strongest upper limit is obtained for \(k=1\) near a characteristic mass of \(3\times 10^7\,{\rm g}\) and are approximately \(\fpbh\simeq 10^{-9}\) for the monochromatic population and \(\fpbh\simeq9\times10^{-11}\) for the log-normal population with \(\sigma=0.5\).

The log-normal calculation shows that the low-mass tail of an extended distribution can substantially strengthen the neutrino limits. At equal characteristic mass, \(M_0 = M_c\), PBHs in the low-mass tail are lighter and therefore radiate more efficiently than PBHs near the median mass. For the mass ranges considered in this work, the ratio of the strongest
log-normal and monochromatic upper limits reaches values as small as
\(\mathcal{R}_{90}\sim10^{-6}\) for \(\sigma=0.5\) and
\(\mathcal{R}_{90}\sim10^{-9}\) for \(\sigma=1\), depending on \(k\). Thus, within the adopted memory-burden prescription, evaluating a monochromatic constraint at the median mass instead of a log-normal distribution will underestimate the strength of the corresponding constraint by several orders of magnitude.

We also examined whether a representative 30-event signal in IceCube-Gen2 radio and GRAND200k could distinguish the two mass-function hypotheses. The clearest discrimination occurs for \(k_{\rm true}=2\). For the log-normal benchmark, the combined dataset gives a median \(\ln B_{\rm LN,mono}\simeq 7.15\), providing very strong support under the local prior. However, when the shared mass prior is broadened, the median decreases to \(1.43\), corresponding to positive evidence, as discussed in Appendix~\ref{app:prior_sensitivity}. For the corresponding monochromatic benchmark, the combined median is \(\ln B_{\rm LN,mono}\simeq-1.94\), giving positive evidence for the correct model. For \(k_{\rm true}=4\), the combined
medians are approximately \(0.54\) and \(1.75\) for the monochromatic and log-normal benchmarks, respectively. The incorrect positive preference in the monochromatic case shows that the generating mass function cannot be identified reliably in this stronger-suppression scenario.

The parameter estimation also showed a strong dependence on the mass-function hypothesis and the suppression strength. For the monochromatic \(k_{\rm true}=2\) benchmark, the \(68\%\) credible intervals
give 
\begin{equation}
\log_{10}\!\left(\frac{M_0}{\mathrm{g}}\right)
=
4.739^{+0.091}_{-0.103},
\qquad
k=2.007^{+0.042}_{-0.039},
\end{equation}
closely recovering the benchmark values. The corresponding log-normal reconstruction is substantially broader, with
\begin{equation}
\begin{aligned}
    \log_{10}\!\left(\frac{M_c}{\mathrm{g}}\right) &=
    5.393^{+0.683}_{-0.638}, \\
    k &=  2.088^{+0.240}_{-0.178}, \\
    \sigma &= 1.141^{+0.256}_{-0.355},
\end{aligned}
\end{equation}
while the PBH abundance remains poorly reconstructed for both hypotheses. The \(k_{\rm true}=4\) reconstructions are broader in comparison, indicating
stronger degeneracies between the parameters involved. These Bayesian results are conditional on the adopted 30-event benchmarks and local prior ranges.

Overall, considering an extended mass function has two significant consequences. The upside of a wider mass function is the high energy tail, which causes a drastic enhancement of the overall flux resulting in a significantly increased discovery potential for smaller PBH populations. However, a downside comes in the need for additional parameterization to describe the mass distribution itself. This adds further parameter degeneracies that considerably weaken the constraints on the other theory parameters \(M\) and \(k\).

We emphasize that neither the monochromatic nor the log-normal mass distributions are meant to correspond to any specific model of PBH formation. The virtue of studying these two cases in particular is the following. First, the monochromatic distribution, although rather unrealistic, is often taken as a simplifying assumption in various studies of memory-burdened PBHs. Our results thus highlight the specific cautions that should be taken when interpreting the results of such studies. Second, these two distributions serve as two edge cases outlining the key differences between a very broad and very narrow mass distribution. The fact that even these two edge cases can be exceedingly hard to distinguish, as shown in this work due to the existence of significant parameter degeneracies, highlights an important challenge in pinning down the model behind a future PBH signal.

In conclusion, high- and ultra-high-energy neutrino events provide a direct probe of light PBHs whose survival is enabled by the memory-burden effect. The initial PBH mass distribution is central to the neutrino signal because it can alter abundance limits by several orders of magnitude and determine whether the population model and physical parameters can be reconstructed reliably.

\acknowledgments
This work is supported by the Swedish Research Council (Vetenskapsrådet) through grant 2023-05141.

\appendix 

\section{Bayesian priors and sampling settings}
\label{app:priors}

The priors used for the Bayesian model comparison and parameter reconstruction are summarized in Tables \ref{tab:priors_model_comparison} and \ref{tab:priors_reconstruction}. These priors are intended for forecasts of detectable PBH signals, not for a global scan of the PBH parameter space. For model comparison, the monochromatic and log-normal hypotheses are fitted over common mass and abundance prior ranges for each value of \(k_{\rm true}\). For parameter reconstruction, the priors are centered on the benchmark point of the model being reconstructed.

For the Bayesian model comparison, we use a minimum of 300 live points, a requested minimum effective sample size (ESS) of 300, and a remaining-evidence stopping tolerance of \(d\ln Z=0.4\). A fit is accepted when ESS \(\geq 300\) and \(\sigma_{\ln Z} \leq 0.40\). Fits that do not satisfy these criteria are rerun for the same simulated dataset and priors using at least 600 live points, a requested minimum ESS of 600, and \(d\ln Z = 0.2\). All 50 realizations entering each reported summary satisfy the criteria. For the parameter reconstruction analysis, we use a minimum of 1000 live points, a requested minimum effective sample size of 1000, and a stopping tolerance of \(d\ln Z=0.3\). A reconstruction run is included in the reported posterior summary only if it satisfies \({\rm ESS}\geq1000\) and \(\sigma_{\ln Z}\leq0.20\).

\begin{table*}[t]
\centering
\caption{Prior ranges used for Bayesian model comparison. The monochromatic and log-normal hypotheses use common mass and abundance prior ranges for each value of \(k_{\rm true}\).}
\label{tab:priors_model_comparison}
\small
\setlength{\tabcolsep}{5pt}
\begin{ruledtabular}
\begin{tabular}{@{}lccccc@{}}
Mass function & \(k_{\rm true}\) & Mass prior & \(\log_{10}\fpbh\) & \(k\) & \(\sigma\) \\
\hline
Monochromatic
& \(2\)
& \(\log_{10}(M_0/{\rm g})\in[4.0,\,7.0]\)
& \([-7.63,\,-3.62]\)
& \([1.5,\,2.5]\)
& -- \\
Log-normal
& \(2\)
& \(\log_{10}(M_c/{\rm g})\in[4.0,\,7.0]\)
& \([-7.63,\,-3.62]\)
& \([1.5,\,2.5]\)
& \([0.5,\,1.5]\) \\
Monochromatic
& \(4\)
& \(\log_{10}(M_0/{\rm g})\in[0.85,\,3.70]\)
& \([-6.92,\,-2.93]\)
& \([3.5,\,4.5]\)
& -- \\
Log-normal
& \(4\)
& \(\log_{10}(M_c/{\rm g})\in[0.85,\,3.70]\)
& \([-6.92,\,-2.93]\)
& \([3.5,\,4.5]\)
& \([0.5,\,1.5]\) \\
\end{tabular}
\end{ruledtabular}
\end{table*}

\begin{table*}[t]
\centering
\caption{Prior ranges used for parameter reconstruction. The mass and
abundance priors are model-specific and centered on the benchmark point being
reconstructed.}
\label{tab:priors_reconstruction}
\small
\setlength{\tabcolsep}{5pt}
\begin{ruledtabular}
\begin{tabular}{@{}lccccc@{}}
Mass function & \(k_{\rm true}\) & Mass prior & \(\log_{10}\fpbh\) & \(k\) & \(\sigma\) \\
\hline
Monochromatic
& \(2\)
& \(\log_{10}(M_0/{\rm g})\in[4.0,\,5.5]\)
& \([-4.62,\,-3.62]\)
& \([1.5,\,2.5]\)
& -- \\
Log-normal
& \(2\)
& \(\log_{10}(M_c/{\rm g})\in[4.0,\,7.0]\)
& \([-7.63,\,-6.63]\)
& \([1.5,\,2.5]\)
& \([0.5,\,1.5]\) \\
Monochromatic
& \(4\)
& \(\log_{10}(M_0/{\rm g})\in[0.85,\,1.70]\)
& \([-3.93,\,-2.93]\)
& \([3.5,\,4.5]\)
& -- \\
Log-normal
& \(4\)
& \(\log_{10}(M_c/{\rm g})\in[0.85,\,3.70]\)
& \([-6.92,\,-5.92]\)
& \([3.5,\,4.5]\)
& \([0.5,\,1.5]\) \\
\end{tabular}
\end{ruledtabular}
\end{table*}

\section{Parameter reconstruction for \(k_{\rm true}=4\)} 
\label{app:k4-reconstruction}

\begin{figure*}[t] 
    \centering 
    \includegraphics[width=\textwidth] {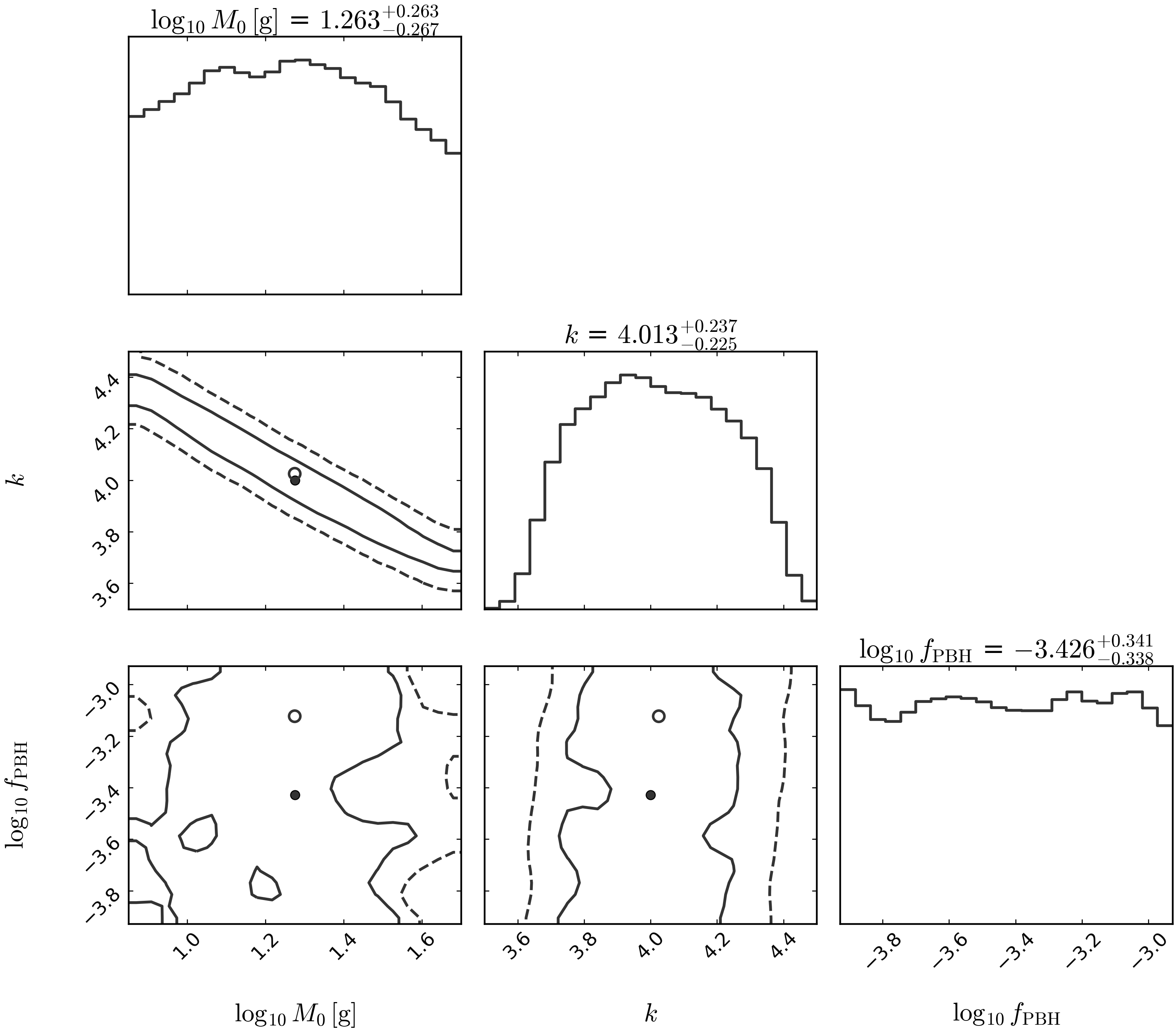} 
    \caption{Posterior reconstruction for the monochromatic benchmark with \(k_{\rm true}=4\), \(M_0^{\rm true}=1.88\times10^{1}\,\mathrm{g}\), and \(\fpbh^{\rm true}=3.73\times10^{-4}\), using the combined IceCube-Gen2 radio and GRAND200k expected-count dataset containing 30 signal events. The plotting conventions are the same as in Fig. \ref{fig:mono-k2-reconstruction}.}
    \label{fig:mono-k4-reconstruction} 
\end{figure*}

\begin{figure*}[t] 
    \centering \includegraphics[width=\textwidth] {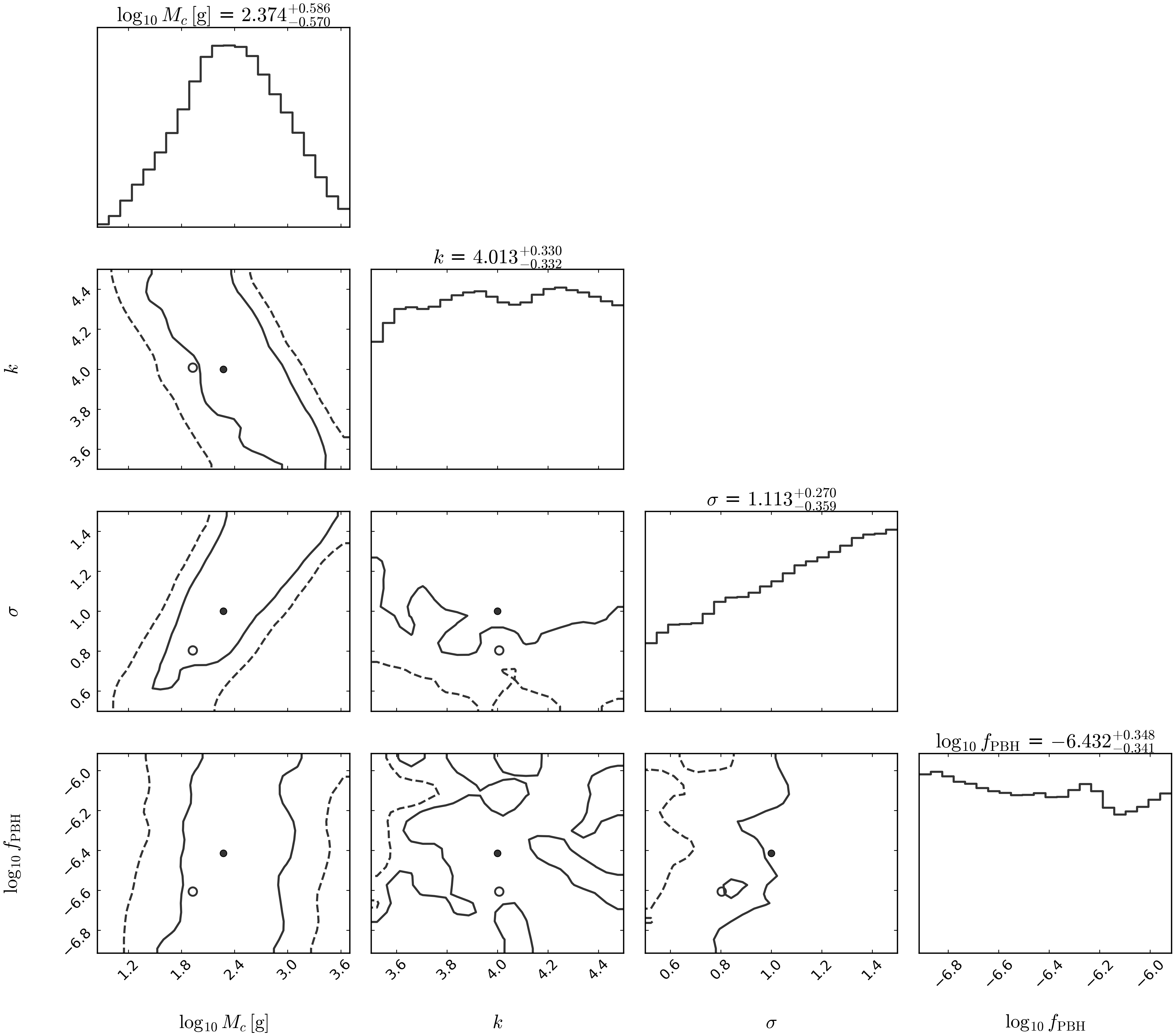} 
    \caption{Posterior reconstruction for the log-normal benchmark with \(k_{\rm true}=4\), \(M_c^{\rm true}=1.88\times10^{2}\,\mathrm{g}\), \(\sigma_{\rm true}=1\), and \(\fpbh^{\rm true}=3.85\times10^{-7}\), using the combined IceCube-Gen2 radio and GRAND200k expected-count dataset containing 30 signal events. The plotting conventions are the same as in Fig. \ref{fig:mono-k2-reconstruction}, with the addition of the log-normal width parameter \(\sigma\).} 
    \label{fig:lognormal-k4-reconstruction} 
\end{figure*}

The parameter reconstructions for the \(k_{\rm true}=4\) benchmarks are shown in Figs. \ref{fig:mono-k4-reconstruction} and \ref{fig:lognormal-k4-reconstruction}. These results are included as supporting material because the \(k_{\rm true} = 4\) benchmarks show larger parameter degeneracies than the \(k_{\rm true}=2\) benchmarks.

The \(k_{\rm true}=4\) reconstructions are substantially broader than their \(k_{\rm true}=2\) counterparts. For the monochromatic benchmark, the posterior contains a broad anticorrelated region in \(M_0\) and \(k\), indicating that changes in the initial mass and memory-burden strength can produce similar event
spectra. In the log-normal case, this degeneracy is increased by correlations with the width \(\sigma\), resulting in the weakest reconstruction among the four benchmarks. The abundance parameter remains weakly localized in both cases because it primarily controls the signal normalization and can trade against the other parameters. These results are benchmark- and prior-specific and show that a 30-event signal would not generally provide precise parameter measurements in the stronger-suppression case.

\section{Prior sensitivity}
\label{app:prior_sensitivity}

Bayesian evidences and marginalized parameter posteriors are conditional on the adopted prior ranges. We therefore examine whether the main conclusions change when these ranges are broadened. Across the tested prior variations, the evidence conclusions are broadly stable, with the exception of the \(k_{\rm true} = 2\) log-normal benchmark under a widened mass prior, which substantially weakens the evidence. The mass prior is extended from \(\log_{10}(M/{\rm g})\in[4,7]\) to \([3,8]\), while the remaining priors and simulated datasets are kept fixed. The comparison uses the same 50 Poisson realizations as the baseline analysis.

For the combined dataset, the median decreases from \(\ln B_{\rm LN,mono}=7.15^{+2.93}_{-3.40}\) under the baseline prior to \(1.43^{+0.63}_{-0.49}\) under the wider mass prior. This reduction is consistent with the wider interval allowing the monochromatic hypothesis to access lower masses that better reproduce the high-energy contribution from the log-normal low-mass tail. The preference for the log-normal hypothesis therefore remains positive, but decreases from very strong evidence under the baseline prior to positive evidence under the wider prior. The strength of this result should consequently be interpreted conditionally on the local mass prior adopted in the baseline analysis.

We also repeat the four expected-count parameter reconstructions after widening the prior ranges. For the \(k_{\rm true}=2\) benchmarks, we use
\[
\log_{10}(M/{\rm g})\in[3,8],
\qquad
k\in[1,3],
\]
while for \(k_{\rm true}=4\) we use
\[
\log_{10}(M/{\rm g})\in[0.5,4.5],
\qquad
k\in[3,5].
\]
For the log-normal model, the width prior is extended from \(\sigma\in[0.5,1.5]\) to \([0.5,2]\), and the PBH abundance prior is widened from \(\pm0.5\) to \(\pm1\) dex around its benchmark value.

For both monochromatic benchmarks, the posterior medians of \(M_0\) and \(k\) remain essentially unchanged. Their credible intervals are also stable for \(k_{\rm true}=2\), while those for \(k_{\rm true}=4\) broaden under the enlarged prior volume. The
log-normal posteriors show a stronger prior dependence, broadening and shifting along the degeneracy directions involving \(M_c\), \(k\), and \(\sigma\). The benchmark parameter values nevertheless remain within the corresponding \(68\%\) credible intervals in all four reconstructions. The log-normal parameter constraints should therefore be interpreted conditionally on the adopted local prior ranges.
\clearpage
\twocolumngrid
\bibliography{refs}

\end{document}